\documentclass[12pt,preprint]{aastex}

\def\kms{km~s$^{-1}$}
\def\degpnt{^{\circ}\kern-1.7mm.\kern+.35mm}
\def\arcpnt{"\kern-1.7mm.\kern+.35mm}
\def\minpnt{'\kern-1.0mm.\kern+.30mm}

\slugcomment{To appear in AJ}

\shorttitle{Novae in M81}
\shortauthors{Neill \& Shara}

\begin{document}


\title{The H$\alpha$ Light Curves and Spatial Distribution of Novae in M81}


\author{James D. Neill\altaffilmark{1}}
\affil{Astronomy Department, Columbia University, New York, NY 10025}
\email{neill@astro.columbia.edu}

\author{Michael M. Shara}
\affil{American Museum of Natural History, 79th and Central Park West
	New York, NY, 10025}
\email{mshara@amnh.org}


\altaffiltext{1}{Graduate Research Fellow, American Museum of Natural
History}


\begin{abstract}

We present the results of a preliminary H$\alpha$ survey of M81 for
novae conducted over a 5 month interval using the 5' field of view
camera (WFCAM) on the Calypso Telescope at Kitt Peak, AZ.  We observed
M81 nearly every clear night during this interval, covering the entire
galaxy, and discovered 12 novae.  Our comprehensive time coverage
allowed us to produce the most complete set of H$\alpha$ light curves
for novae in M81 to date.  A raw nova rate for M81 gives 23 yr$^{-1}$
which, because of the nature of our survey, is a hard lower limit.
An analysis of the completeness in our survey gives a corrected nova
rate of 30 yr$^{-1}$.  This agrees well with the rate of 
33$^{+13}_{-8}$ yr$^{-1}$, derived from Monte Carlo simulations using 
nova light curves and survey frame limits.  The spatial distribution 
of the novae we discovered follows the bulge light much better than the
disk or total light according to Kolmogorov - Smirnov tests of their radial
distributions.  The asymmetry in the distribution of novae across the major 
axis line of M81 implies a bulge-to-disk nova ratio of $>$ 9 and 
supports the idea that novae originate primarily in older stellar 
populations.

\end{abstract}




\section{Introduction}

Extragalactic novae are particularly appealing objects for studies
of close binary evolution.  They are {\it the} tracer of interacting
close binary stars, visible to much greater distances than any other
well studied standard candle except supernovae (novae display M$_v$
up to $-10$).  By studying the novae in a nearby galaxy, one can gather
a homogeneous sample of objects, all at the same distance, that is not
plagued by the selection effects hampering an analysis of the cataclysmic
variable (CV) population in our own galaxy.

Given the large fraction of all stars that exist in binaries, the
effort to understand binary formation and evolution has wide reaching
implications for understanding stellar populations.  There are currently
many unsolved problems in the theory of close binary formation and
evolution that are difficult to tackle using CVs in our own galaxy
because of the selection effects mentioned above.

The prediction that the nova rate should correlate with the star formation
history in the underlying stellar population \citep{yun97} is one example.
This prediction is based on our understanding of how close binaries
form and evolve, combined with our understanding of how the mass of a
nova progenitor influences the nova outburst.  Massive white dwarfs
come from massive progenitors.  These massive white dwarfs erupt as
novae frequently as they need only accrete a small amount of hydrogen
from their companions to explode as novae.  This, in turn, implies
that stellar populations with a low star formation rate should have
a correspondingly low nova rate, i.e., early-type galaxies with older
stellar populations should have a lower luminosity specific nova rate
(LSNR) than late-type galaxies with ongoing star formation.

Efforts have been made to detect a trend in LSNR with galaxy type
\citep{del94,sha00}, but the random errors are too large to date for a
meaningful comparison.  Typical extragalactic nova surveys are carried out
using short runs, and significant assumptions must be made about the mean
lifetimes of novae in order to derive nova rates \citep{sha01}.  It is
also rare that an extragalactic survey has been able to spatially cover an
entire galaxy and avoid making some assumption about the distribution of
novae with light to derive a galactic nova rate.  The resulting errors in
the nova rates are large, but systematic biases are far more pernicious.

In order to accurately compare nova rates with the underlying stellar
population, we are pursuing a research program that uses the comprehensive
time and spatial coverage afforded by a dedicated observatory.  We have
begun our program with M81 and used 5 continuous months of observing time
to produce a survey that requires no assumptions about nova distribution
or mean lifetime.  Not since \citet{arp56} surveyed the central parts
of M31 has such extensive, continuous coverage of a galaxy for novae
been attempted.

Our survey updates the 5 year photographic survey of M81 for novae
reported by \citet{sha99}.  They found 23 novae, evenly divided over the
disk and central bulge.  Significant incompleteness must be present in
that survey due to photographic saturation in the bright central regions
of the galaxy and due to large gaps in their time coverage.  The results
we present below show that a comprehensive spatial and temporal survey
is required to minimize the systematic effects of incompleteness.

\section{Observations}

\begin{figure}
\plotone{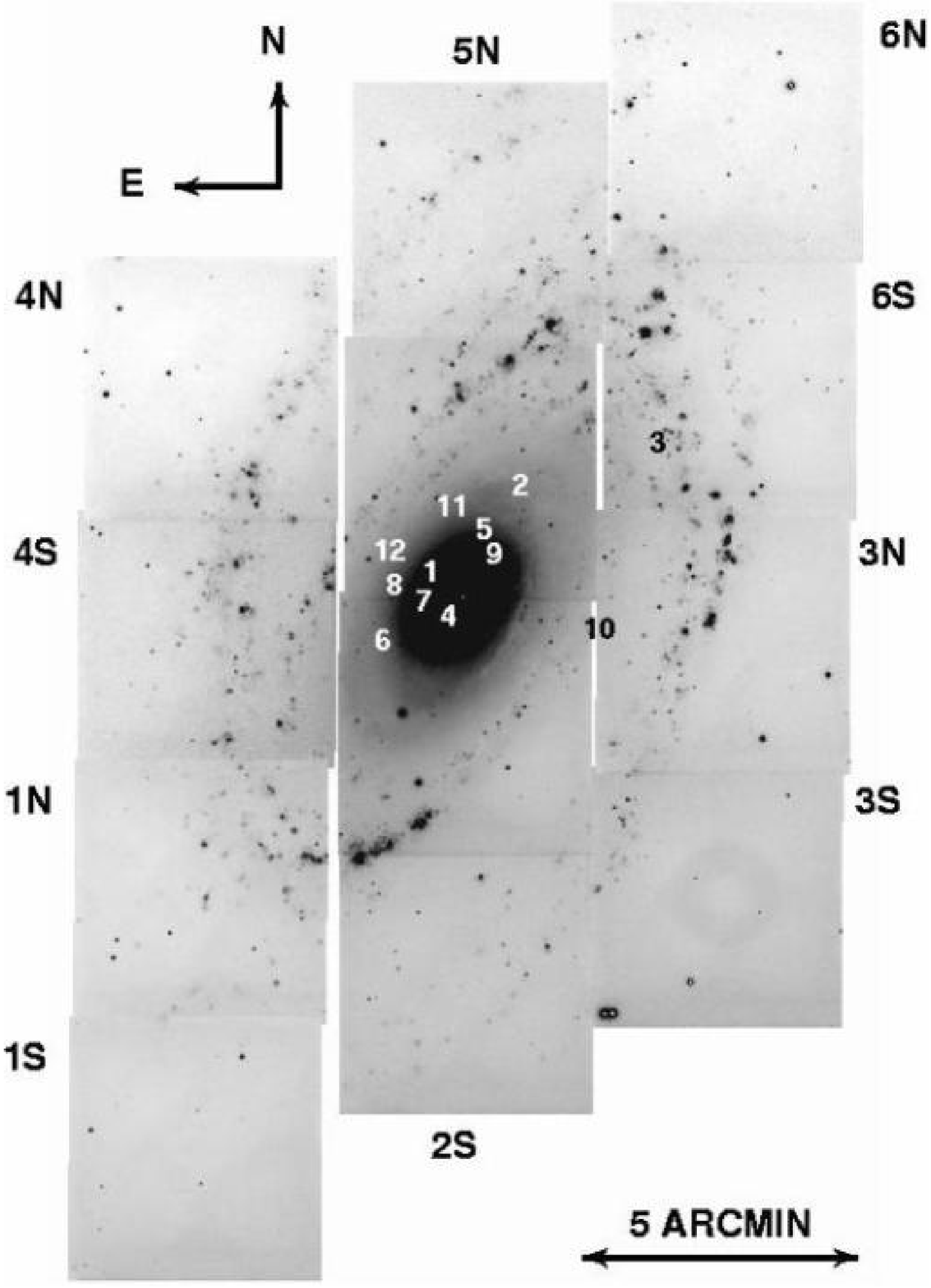}
\caption{H$\alpha$ mosaic of the 12 M81 survey fields (labeled)
with the positions of the 12 novae discovered in the 2003 season indicated 
as white or black numbers.  Note that most of the novae are close to the
nucleus of the galaxy and the asymmetry of the nova distribution across
the major axis.
\label{fig_pos}
}
\end{figure}

We used the WFCAM 2048x2048 pixel CCD on the Calypso 1.2m telescope at
4x4 binning for our observations of M81.  This configuration yields a
pixel scale of 0\arcpnt6 px$^{-1}$ and a field size of 5' on a side.
The seeing for our observations had a median of 1\arcpnt5 and ranged
from 1\arcpnt0 to 2\arcpnt5.  In an effort to cover the entire spatial
extent of M81 in our search for novae we divided M81 into 12 fields
covering roughly 15' in Right Ascension and 23' in Declination.
Figure~\ref{fig_pos} is an H$\alpha$ mosaic of the 12 fields showing
the extent of our spatial coverage, the identification of the fields,
and the location of the 12 novae discovered.

\begin{figure}
\plotone{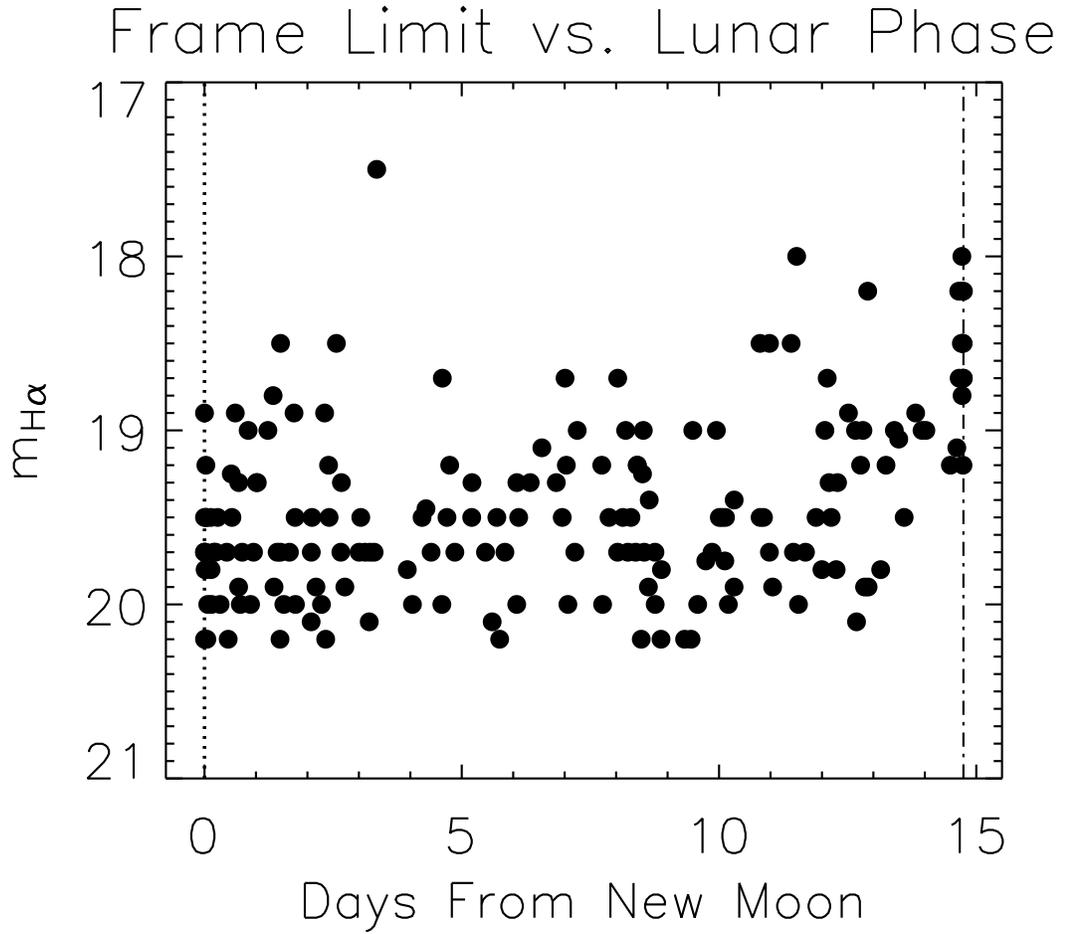}
\caption{Frame limits in $m_{H\alpha}$ versus days from the New Moon.  The
dotted line indicates the New Moon, while the dot-dashed line indicates the
Full Moon.  The moon's impact on the survey depth is limited to the three
days centered on Full Moon and has a maximum amplitude of less than a
magnitude.  Since novae last longer in H$\alpha$, the chance that we
missed a nova because of the moon is very small.
\label{fig_moonph}
}
\end{figure}

All observations were taken through an H$\alpha$ filter with a
30\AA~full width at half-maximum (FWHM) bandpass.  This filter was
chosen for several reasons.  Because of the longer duration of novae
in H$\alpha$ compared to the B-band \citep{cia87}, we minimize the
possibility of missing novae due to inevitable gaps in coverage.
The redder wavelengths observed with the H$\alpha$ filter means that
our images are less influenced by internal extinction in M81, and less
influenced by scattered moonlight during the fuller phases of the moon.
To illustrate this point, Figure~\ref{fig_moonph} shows the distribution
of frame limits (described in \S~\ref{novdet}) with days from New Moon.
Only within 1.5 days of the full moon is any effect seen.  It is also
our goal to explore the H$\alpha$ light curve as a tool for understanding
the physics of nova outbursts.  So far attempts to do this have failed,
but with our dense time sampling the possibility opens up that features
of the H$\alpha$ rise, previously poorly observed, may correlate with
properties of the nova progenitor.

Each individual exposure was 1200s and we attempted to get at least 5
exposures on a given field in one night for a total exposure time of
100 minutes per epoch.  Although most of our epochs reached this goal,
observing conditions varied considerably and so the number of useful
exposures per epoch ranges from 2 to 9.

Ideally we would have observed all 12 fields every clear night.  This was
impossible to achieve with the requirement that we get 100 minutes of
exposure per epoch.  Thus, we cycled through the 12 fields over the
course of a varying number of nights depending on the time available on
a given night subject to weather, M81's availability, and occasional
equipment problems.  Guide stars were not readily available for some
of the fields or were faint, with the result that the number of usable
survey epochs for each field varies from 5 to 21.

Our survey ran from December 31, 2002 (JD 2452638.8) to June 6, 2003
(JD 2452796.6) covering a total of 158.8 days.  At the end of this time,
two novae were still observable in decline in one of the fields (5S).
An additional 15 epochs were obtained of this field to follow the
declining novae until June 26, 2003 (JD 2452816.7), during which time
no new novae were discovered in this field. This period from June 6 to
June 26, 2003 is not included in our nova rate calculations.

Table~\ref{tab_obs} summarizes the observations in our survey.  The field
designations correspond to those in Figure~\ref{fig_pos}, with the
exception that field 5S and 2N are unlabeled (but their location is
obvious).  The fields with Bulge in the last column also contain a
significant amount of M81's disk as can be seen from Figure~\ref{fig_pos}.

\section{Reduction}

All exposures were bias subtracted and flat fielded using the standard
tools in IRAF \citep{tod86}.  The exposures for a given epoch (from 2
to 9) were then registered and combined to produce a coadded image for
each epoch using the following DAOPHOT programs \citep{ste87}: DAOPHOT to
measure the point sources, DAOMATCH and DAOMASTER to derive and refine the
transformations, and MONTAGE2 to perform the registration and coaddition.
The registration master was chosen to be the best coadded image from
the entire set for a given field based on measurements of the seeing
in each coadded image over the entire survey.  The coaddition process
removed all but a few cosmic rays.

\section{Nova Detection \label{novdet}}

The coadded images were blinked against each other to detect changing
point sources.  Nova candidates were required to be observed in at
least two epochs {\it and} to be missing on an epoch of sufficient
depth coverage to confirm its transient nature.  We also used the raw
images for an epoch to confirm the presence of the candidate in each
individual frame.

For the bulge region of M81 (Fields 2N and 5S), where the intensity
gradient makes detection more difficult, we used the following technique.
We produced a spatial median filtered image of the coadded frame using
a box size of 11 by 11 pixels to remove point sources and preserve the
low frequency structure of the image.  This was then subtracted from the
original coadded frame which removed the intensity gradient of the bulge,
but preserved the point sources embedded within it.  Residual light in the
subtracted image within 20" of the nucleus made detection of the innermost
novae very difficult.  This subtraction technique was performed on each
coadded image of the bulge fields after which they were blinked against
each other.  Since most of the novae were detected in the bulge region,
the subtraction proved to be important for detecting novae in M81.

For each coadded image of each field we determined the frame limit
by using artificial stars and the exact techniques outlined above for
detecting the novae.  Closer to the nucleus of M81, frame limits were
derived from the flattened images.  For field 2N, an additional set of
frame limits was derived near the position of Nova 4.  For field 5S,
two additional sets of frame limits were derived near the positions of
Nova 1 and Nova 5.

Table~\ref{tab_nov_pos} summarizes the 12 novae discovered with our
survey.  Astrometry was derived for each nova using the WCSTOOLS package
\citep{min02} and the Guide Star Catalog-II\footnote{ The Guide Star
Catalog-II is a joint project of the Space Telescope Science Institute and
the Osservatorio Astronomico di Torino. Space Telescope Science Institute
is operated by the Association of Universities for Research in Astronomy,
for the National Aeronautics and Space Administration under contract
NAS5-26555. The participation of the Osservatorio Astronomico di Torino
is supported by the Italian Council for Research in Astronomy. Additional
support is provided by European Southern Observatory, Space Telescope
European Coordinating Facility, the International GEMINI project and the
European Space Agency Astrophysics Division.}.  Positions are accurate
to better than 2" based on the fit residuals to the GSC-II stars.
The corrected nuclear distances were calculated using the equations and
parameters given in \citet{sha99}.

Could our candidates be anything other than novae?  Their positions
in M81 rule out foreground or background objects.  We checked the
position of each candidate against the list of known Hubble-Sandage (HS)
variables in M81 \citep{san84} and found no coincidence.  These bright
blue HS variables are the only known non-nova variables that approach the
brightnesses of our candidates.  In fact, the brightest B-band magnitude
for the HS variables in M81 is 19.1 \citep{san84}.  These are massive
stars and are likely to be much fainter in the red at the H$\alpha$
passband than in the B-band.  Considering the peak magnitudes of our
candidates (see below), there is little chance that any of our nova
candidates is actually one of these bright blue variables\ldots or
anything other than a nova.

\section{Nova Photometry}

Since crowding was not an issue, aperture photometry was used to measure
point source brightnesses in each coadded image.  We used the IRAF {\it
apphot} package to measure point source brightnesses.  Variable seeing was
accounted for by setting the measurement aperture radius in each image to
1/2 the FWHM of the stellar profile to maximize the signal-to-noise ratio.
The FWHM was measured from a set of well isolated stars in each image.

The photometric calibration was achieved using a two-step process, which
was required because of the negligible overlap between fields, and the
lack of H$\alpha$ standards in many of the fields.  The first step was
to tie all the frames to a common system.  This was achieved using the
R-band photometry of \citet{per95}.  An offset from our instrumental
H$\alpha$ magnitudes to the calibrated R magnitudes was calculated for
each epoch of each field using from 18 to 70 stellar sources per field
with an accuracy of better than 0.1 magnitude.  Non-stellar sources (HII
regions) were excluded by virtue of their much brighter instrumental
H$\alpha$ magnitude relative to R than the stellar sources.

Once all magnitudes were on this common R system, an offset was needed
to the standard AB system where $m_{H\alpha} = 0.0$ for $f_{\lambda} =
2.53 \times 10^{-9}$ erg cm$^{-2}$ s$^{-1}$ \AA$^{-1}$.  For our filter
with a FWHM of 30\AA, this gives a zero point flux of $7.59 \times
10^{-8}$ erg cm$^{-2}$ s$^{-1}$.  We used the foreground extinction
corrected and standardized flux measurements of M81 planetary nebulae
(PNe) published by \citet{mag01} to calculate this offset.  They used
a 90\AA~FWHM H$\alpha$ filter which would include [NII] if present.
Our filter excludes [NII] so we had to assume that for the set of PNe
used, the [NII] fluxes are small.  Using our filter zero point flux, we
converted the fluxes from \citet{mag01} into H$\alpha$ filter magnitudes
for the 107 PNe we were able to measure in 7 of our fields.  We then
compared these H$\alpha$ filter magnitudes with the common R system
magnitudes for the PNe to calculate an offset between the two systems.
We calculated a mean offset from 78 PNe of $-0.26\pm0.13$ magnitudes.
We excluded 29 outliers that showed evidence of stronger [NII] by virtue
of their having a lower instrumental H$\alpha$ magnitude than predicted
from the conversion of the flux to H$\alpha$ filter magnitude.  Without
individual spectra of the PNe, this exclusion is less than perfect and
limited our photometric accuracy.

As a check on this calibration, we found one object which is in both the R
catalog of \citet{per95} and the H$\alpha$ flux catalog of \citet{mag01}.
Object 2111 from \citet{per95} with an R magnitude of 19.64 is object
93 from \citet{mag01} which has an H$\alpha$ magnitude (using the zero
point flux above) of 19.42.  This gives an offset of -0.22, in good
agreement with our calibration.

This calibration should allow comparison of our H$\alpha$ nova light
curves with the M31 H$\alpha$ light curves of \citet{cia90}, with one
important caveat.  Because our filter is 2.5 times narrower, many of
the faster novae with expansion velocities $>$ 685~\kms~have overfilled
our 30\AA\ bandwidth.  The M31 novae can have expansion velocities
up to 1715~\kms~and not overfill the 75\AA~filter bandwidth used by
\citet{cia90}.  Because nova ejection velocity is a function of nova
luminosity \citep{sha81}, a direct comparison of the nova luminosity
distribution between M31 and M81 is precluded for this survey.

Table~\ref{tab_nov_phot} presents our calibrated photometry for each
nova at each observed epoch.  The errors in Column 3 are the 1-$\sigma$
internal photometric errors.  The three points in parenthesis for Nova
6 are unfiltered magnitudes from the KAIT telescope \citep{wei03}.
\citet{fil03} report that the spectrum of this nova exhibited strong
double-peaked hydrogen Balmer emission lines and that many Fe II, Ca II,
and O I lines also appeared in emission, and thereby confirm this object
as a nova.

\section{The H$\alpha$ Light Curves}

\begin{figure}
\plotone{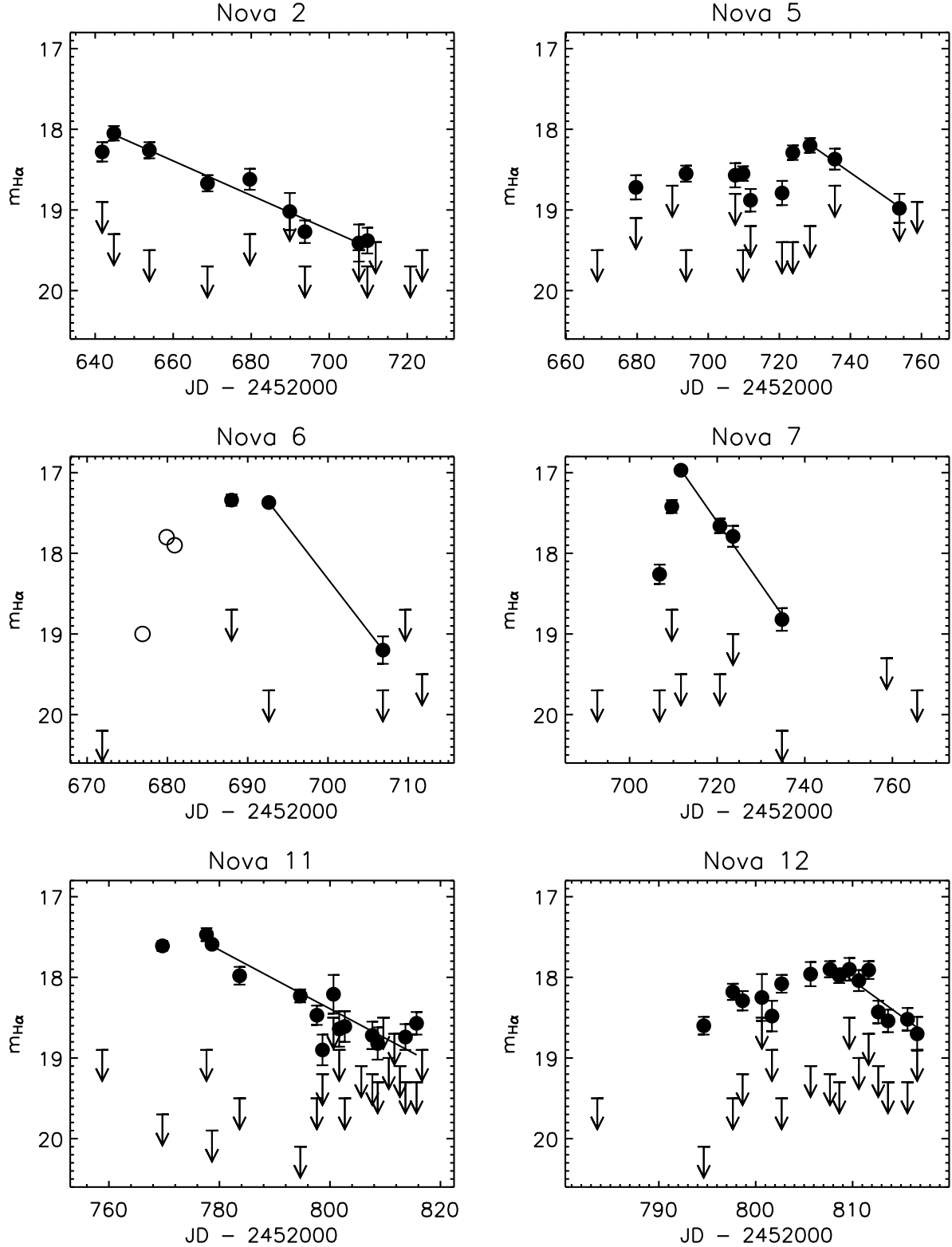}
\caption{The H$\alpha$ light curves of six M81 novae observed to reach maximum
$m_{H\alpha}$.  The solid points are $m_{H\alpha}$ and the frame
limits are indicated by the horizontal line with the downward pointing
arrow.  The open circles for Nova 6 are unfiltered magnitudes from the KAIT
telescope \citep{wei03}.  The thin lines are linear fits to the 
decline rates for each nova.
\label{fig_photmax}
}
\end{figure}

\begin{figure}
\plotone{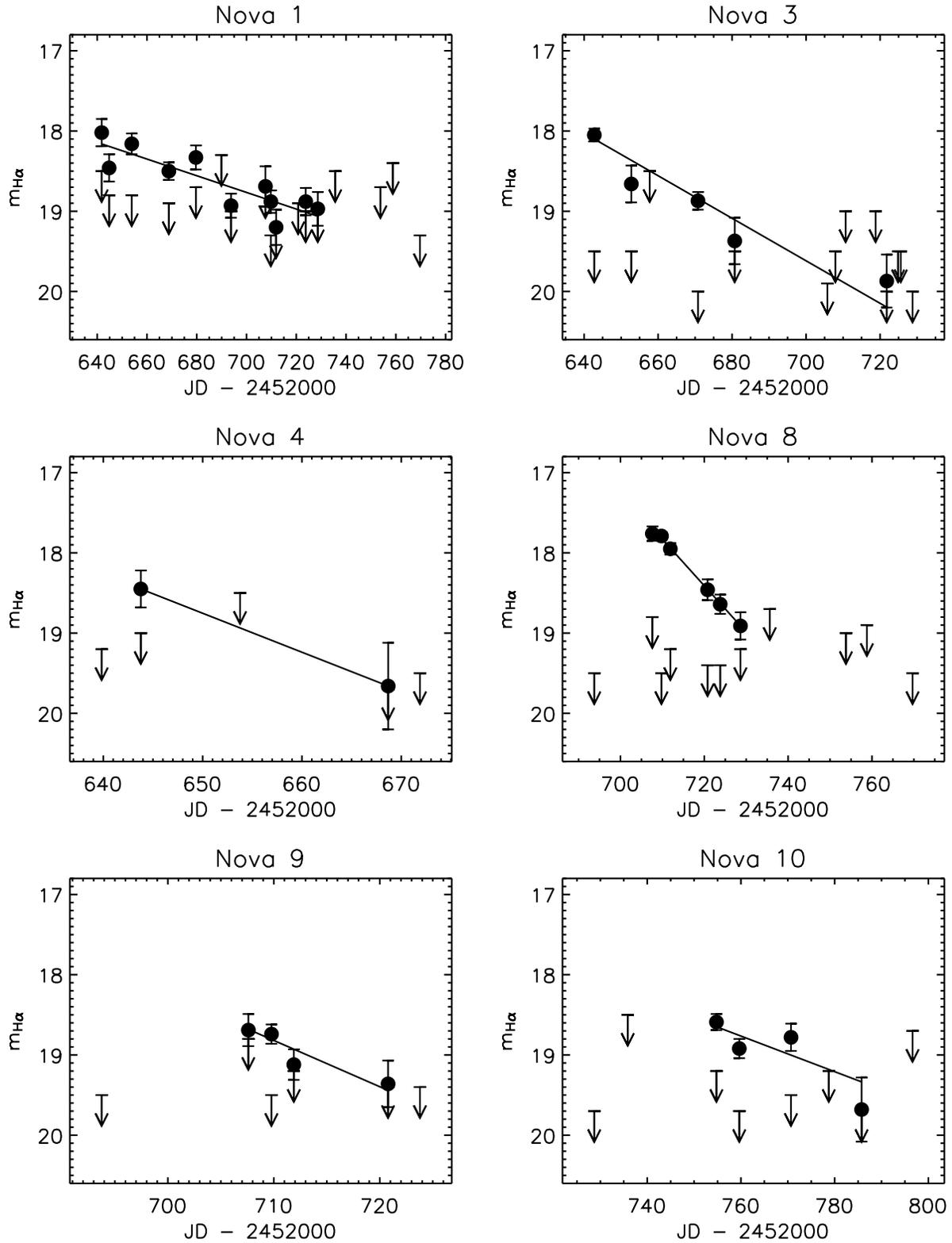}
\caption{As for Figure \ref{fig_photmax}, but for the novae observed days
or weeks after maximum $m_{H\alpha}$.
\label{fig_photdec}
}
\end{figure}

Figure \ref{fig_photmax} and Figure \ref{fig_photdec} present our
calibrated H$\alpha$ light curves for the 12 novae discovered in
M81 during our survey.  Figure~\ref{fig_photmax} presents the 6
novae for which we have observed the maximum $m_{H\alpha}$, while
Figure~\ref{fig_photdec} presents the 6 novae observed days or weeks
after maximum $m_{H\alpha}$.  The frame limits are plotted as short
horizontal lines with downward pointing arrows, while the solid circles
with error bars are the nova observations from Table~\ref{tab_nov_phot}.

A simple linear fit was made to the decline portion of each light curve
to calculate the decline rates in $m_{H\alpha}$~day$^{-1}$.  The thin
lines in Figure~\ref{fig_photmax} and Figure~\ref{fig_photdec} show the
resulting fits.  Table~\ref{tab_nov_prop} presents the properties of the
nova light curves including the rise time and decline rate for each nova.
We consider the H$\alpha$ rise times to be lower limits because we do
not have continuum light curves to to accurately define the outburst time.

\begin{figure}
\plotone{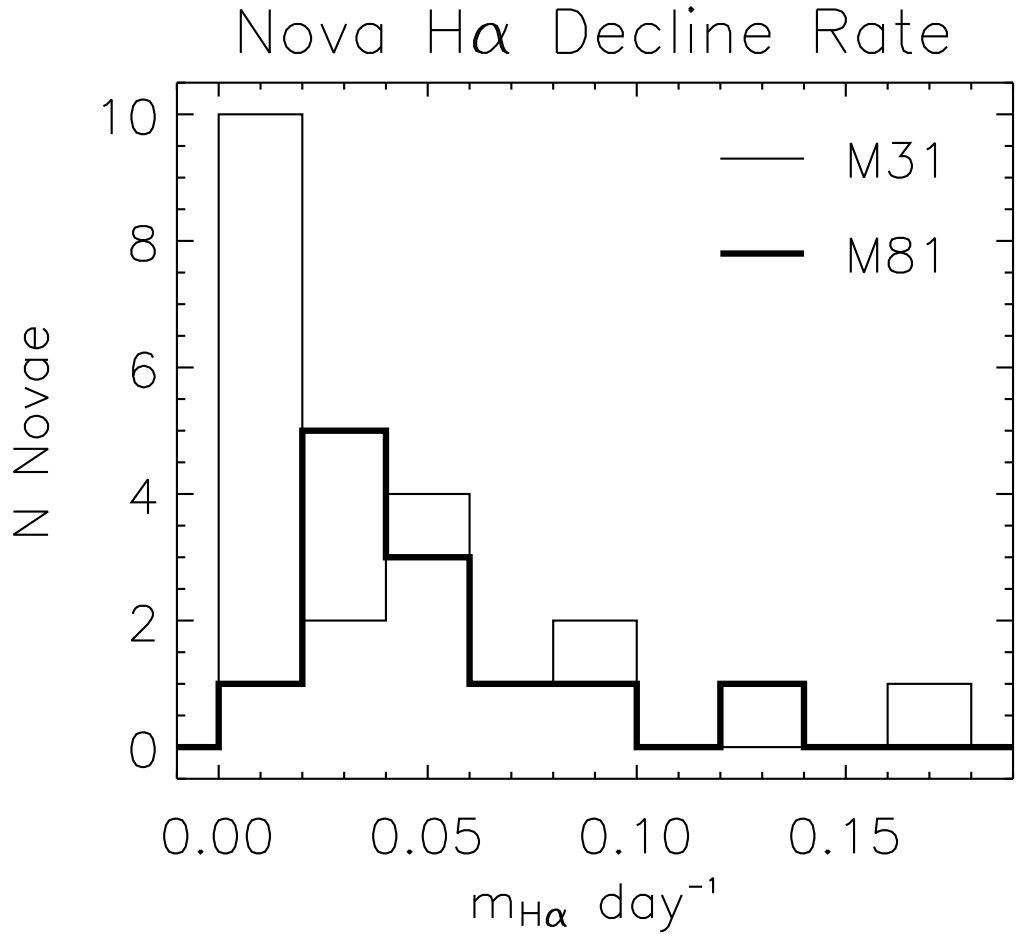}
\caption{A histogram comparing the H$\alpha$ decline rates for our M81 
novae and the
M31 novae of \citet{cia90} and \citet{sha01}.  Note the incompleteness in 
our M81 survey in the slowest bin.  As noted in the text, this is due to the 
5 month overall time baseline of our survey.  The M31 surveys had overall
time baselines of two years or more.
\label{fig_drcomp}
}
\end{figure}

We cannot compare the brightness distribution of our novae to that in M31,
because of the filter-overfilling described above, but we can compare
the nova decline rates.  Figure~\ref{fig_drcomp} shows the comparison of
our decline rates with the M31 decline rates reported in \citet{cia90}
and \citet{sha01}.  We see that the distribution is similar, but there
is an indication of incompleteness in our slowest decline rate bin.
Both M31 surveys spanned at least two years and therefore had longer
overall time baselines.  Our survey is continuous, but only covers a 5
month period.  At the slowest reported decline rate for an M31 nova of
0.0018 $m_{H\alpha}$~day$^{-1}$ \citep{cia90}, in 5 months a nova would
only change by 0.3 magnitudes in H$\alpha$ and would not be seen in our
survey as a transient point source.  In fact, it would take over 500
days before such a slow nova would change by one magnitude.  Based on
Figure~\ref{fig_drcomp}, and accounting for small number fluctuations,
we estimate that there are roughly 10 very slow novae that remain to
be detected in the present survey.  As we extend our survey in time,
we will continue to blink epochs from the season reported here in an
effort to discover these very slowly declining novae.

Two of the light curves show that, in H$\alpha$, novae can take a long
time to reach maximum brightness; Nova 12 took more than 13 days , and
Nova 5 took at least 50 days.  Nova 12 can be compared with Nova 26 in
\citet{cia90} which took $\sim$15 days from outburst to reach maximum.
Our Nova 5 is unprecedented; its 50 day rise is, as far as we know, the
longest ever observed for a nova in H$\alpha$.  Nova 20 from \citet{cia90}
shows a small decline before reaching maximum light, like our Nova 5,
and could have taken a similar length of time to reach maximum.

\begin{figure}
\plotone{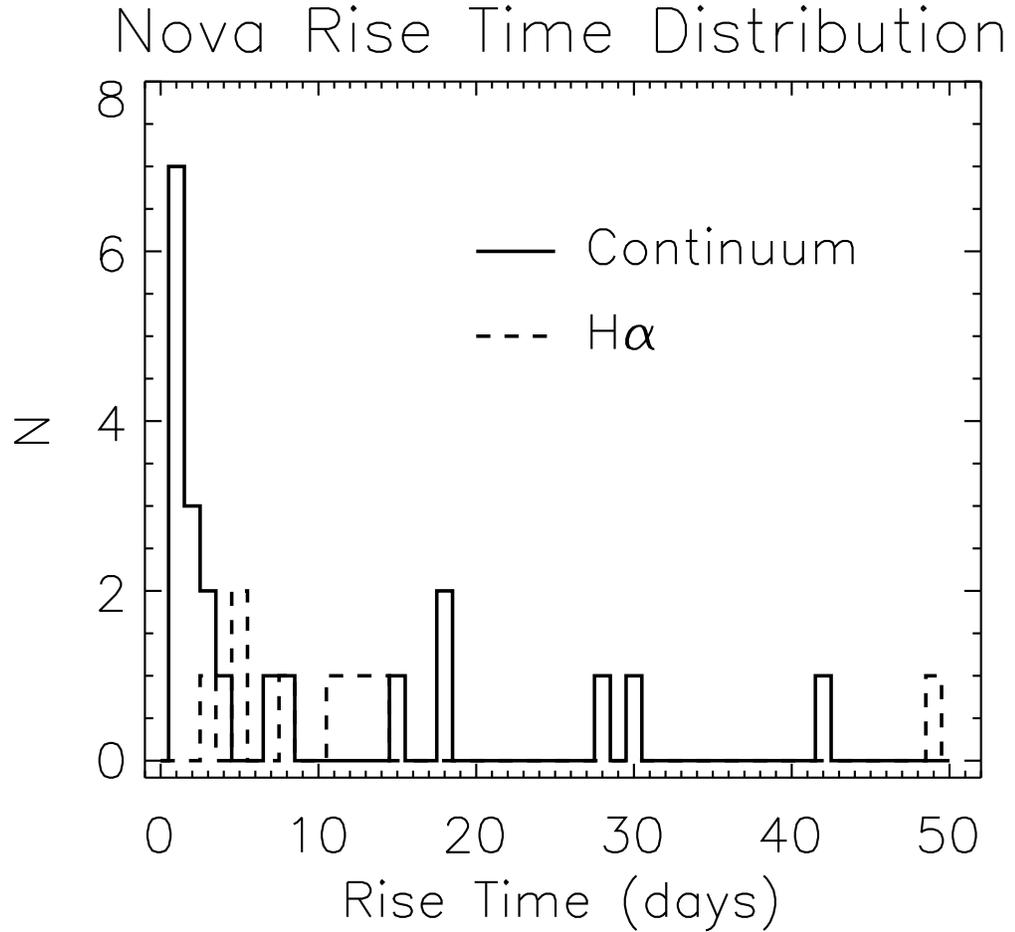}
\caption{A histogram comparing continuum and H$\alpha$ rise times.
The continuum rise times, shown by the solid histogram, are from the 
photographic light curves in \citet{arp56}.  The H$\alpha$ rise times, 
shown by the dashed histogram, are from Table~\ref{tab_nov_prop} and 
from \citet{cia90}.  The arithmetic mean of the continuum rise times 
is 9.0 days, while the arithmetic mean for the H$\alpha$ rise times is 
13.3 days.  Note that the H$\alpha$ rise times presented in this work 
are lower limits.
\label{fig_rise}
}
\end{figure}

Figure~\ref{fig_rise} shows a comparison of nova rise times in H$\alpha$
and the continuum.  The H$\alpha$ data consist of the 6 novae from
this paper (see Table~\ref{tab_nov_prop}) and the three novae with
well observed rises from \citet{cia90}.  The continuum data are from
the photographic light curves presented in \citet{arp56}.  Of the 30
novae discovered by Arp, we used 21 for which we could determine a
reliable rise time.  The continuum distribution has a sharp peak at 1
day, but extends out to nearly as long a rise time as the longest one
for H$\alpha$.  The H$\alpha$ distribution shows no peak, and does seem
to cluster in the range from 3 to 15 days.  The arithmetic mean of the
continuum rise times is 9.0 days.  The arithmetic mean for the H$\alpha$
rise times is 13.3 days, but should only be considered a lower limit given
that all the rise times from Table~\ref{tab_nov_prop} are lower limits.
For the four novae from \citet{cia90} that have both B-band and H$\alpha$
photometry (including Nova Cyg 1975), we see that the H$\alpha$ maximum
always occurs after the B-band maximum.  It appears that H$\alpha$
rise times are generally longer than continuum rise times, but without a
large set of light curves in both continuum and H$\alpha$ for each nova,
we cannot say more.  We hope to remedy this in the next observing season
for M81 (see below).

The slow rise times in H$\alpha$ of some novae may lead to inaccuracy
in the calculation of nova rates if only the maximum light and decline
rates are used, especially with a survey, such as ours, with many closely
spaced epochs.  Nova light curves which cover the rising portion of
the outburst and allow determination of the decline rate are needed for
calculating accurate completenesses and nova rates.

\section{Completeness}

Because of the dense time sampling and variable depth of our survey,
we developed an approach to finding the completeness of each field that
is more appropriate than assigning a single frame limit to the entire
survey.  With a set of relatively complete H$\alpha$ nova light curves
and the frame limit for each epoch of each field, we can calculate the
completeness for a given field by simulating the random outburst times
of a large sample ($10^5$) of novae.  This approach assumes that we
have a set of H$\alpha$ light curves that is representative of the nova
population in M81.

\subsection{Representative H$\alpha$ Light Curves}

Of the 12 light curves reported in this work, three novae have sufficient
coverage of the rising part of the light curve and have reasonably well
determined decline rates to qualify for this set: Novae 5, 7, and 12
(see Figure~\ref{fig_photmax}).  We can combine these with the H$\alpha$
photometry of the 8 novae listed in Table~11 of \citet{sha01} to get
a representative set of 11 H$\alpha$ nova light curves to use in our
random outburst simulation.

In order to include the M31 light curves in our representative set, we
need to correct for the difference in galaxy distances and the difference
in filter bandwidths.  Using the distance moduli quoted in \citet{sha00}
we apply a correction to the M31 novae of +3.48 magnitudes to place
them at the distance of M81.  The filter bandpass we used is 2.5 times
narrower than used by \citet{cia90} for the M31 novae and so they would
appear 1 magnitude brighter in our survey making the correction +2.48
magnitudes.  If the M31 novae had, in fact, been observed with our filter,
they would have overfilled the bandpass by varying amounts and therefore
been fainter by up to a factor of 2.  Spectroscopic observations of M31
novae by \citet{cia83} show typical emission line widths of 40-60~\AA.
However, spectra of novae in M31 reported by \citet{tom92} show the
slowest, and therefore faintest, novae to have expansion velocities
of less than 685~\kms, which would not overfill our filter.  Since we
are interested in calculating the completeness at the faint end, we
elect not to apply an uncertain overfilling factor to the M31 novae.
Therefore, our total correction to the M31 novae to include them with
our M81 photometry through a 30~\AA~filter is +2.48 magnitudes.

To see if this set is truly representative of novae in M81, we can
compare the range of decline rates in this set (0.0044 to 0.087 mag
day$^{-1}$) with Figure~\ref{fig_drcomp}.  We see that this range covers
the majority of the novae in this figure.  By adding the slower novae
from M31, we account for the very slow novae we are apparently missing,
according to Figure~\ref{fig_drcomp}.  We can also compare the range of
maximum magnitudes in this set ($m_{H\alpha}$ from 17.0 to 20.9) with the
maximum magnitudes reported in \citet{cia90} for M31.  After accounting
for the magnitude offset as described above, we see that our set spans
the majority of the range of novae in M31.  For the following analysis,
we will assume that our 11 novae are a representative set for M81.

\subsection{Random Outburst Simulations}

We used each nova in this representative set to make a random outburst
simulation for each set of frame limits we measured. This produced 11
simulations for each of the 15 frame limit sets for a total of 165
simulations.  To create each simulation, which consisted of 10$^5$
trials, we used an uniformly distributed random number generator to
shift the representative nova light curve in time so that its outburst
occurred within the 159 days of our survey.  The frame limit set being
studied was then compared with each shifted light curve to determine
if it would have been a valid nova candidate in that field.  A valid
candidate must pass the criteria describe above: it must be brighter
than at least two frame limits in the set, {\it and} it must be fainter
than at least one of the frame limits, to confirm its transient nature.
The number of times a simulated nova was identified as a candidate was
divided by the number of trials to produce a fractional completeness
for that representative nova in the frame limit set being studied.

In order to compare the randomly shifted nova light curve with the frame
limits in the set, and determine if the simulated nova would have been a
candidate in the field being studied, the following algorithm was adopted.
Any nova magnitude on a frame epoch before outburst was set to 99.9 (i.e.
a non-detection).  Nova magnitudes on frame epochs between two light
curve points were calculated using linear interpolation.  Nova magnitudes
on frame epochs after the last point in the light curve were calculated
using the measured decline rate for the nova.  This is much more accurate
than extrapolating beyond the end of the light curve since the last points
in a nova light curve tend to be the faintest and therefore the noisiest.

\begin{figure}
\plotone{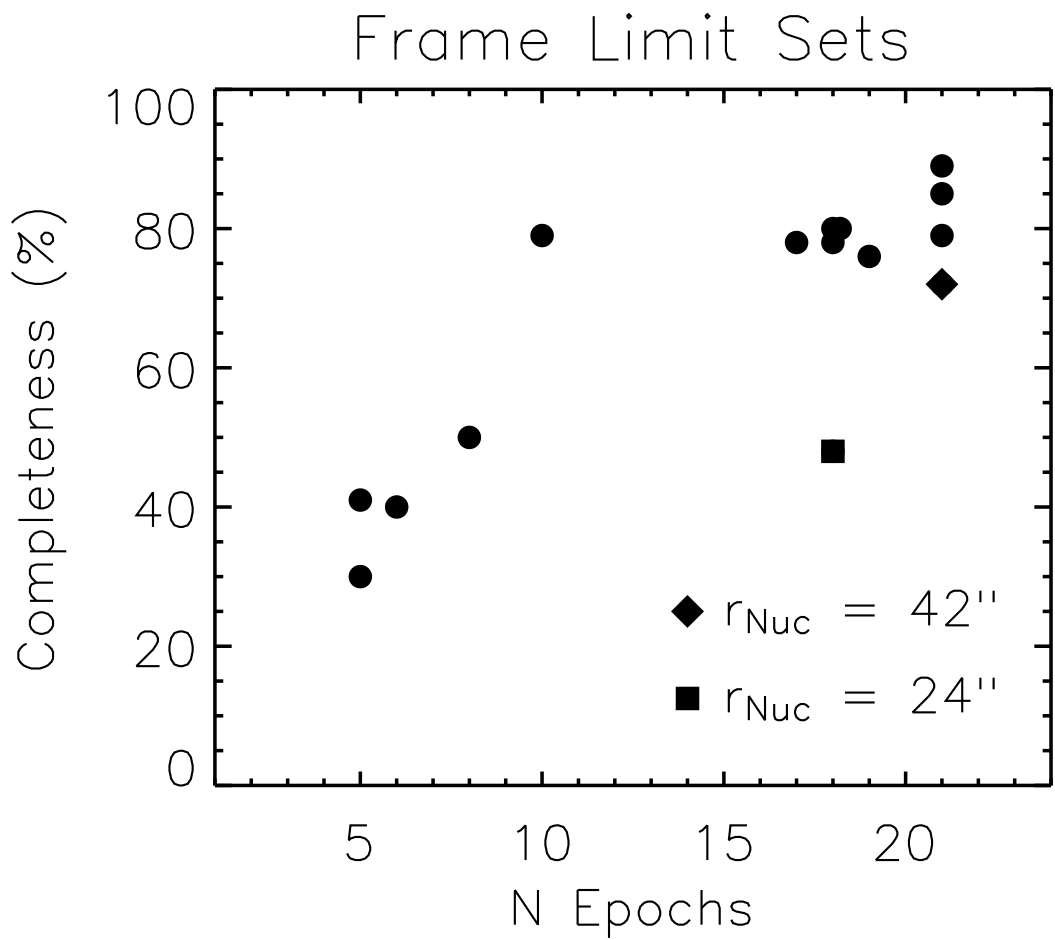}
\caption{The completeness of the 15 frame limit sets as a function of the 
number of epochs from the random outburst simulations.  The trend with the
number of epochs is obvious.  A trend with nuclear distance, r$_{Nuc}$, is
also shown using the two frame limit sets closest to the nucleus of
M81.  For Field 5S, the lowest completeness frame limit set is indicated 
by a diamond.  For Field 2N, the lowest completeness frame limit set is
indicated by a square.  These points illustrate that, in spite of having many
epochs, within an arcminute of the nucleus of M81 a significant 
incompleteness exists.
\label{fig_field_comp}
}
\end{figure}

The completeness as a percentage, averaged over the 11 representative
novae, is given for the 15 frame limit sets in Table~\ref{tab_field_comp}
and plotted in Figure~\ref{fig_field_comp}.  Two correlations are
apparent.  The first and strongest is the correlation with the number
of observed epochs.  The lowest completeness is obtained for the fields
with the smallest number of observed epochs and suggests a substantial
incompleteness for these fields.  The other correlation is with nuclear
distance.  For fields 2N and 5S, the completeness drops considerably as
the distance to the nucleus decreases.  For a frame limit set measured at
42" from the nucleus of M81, and plotted in Figure~\ref{fig_field_comp}
as a filled diamond, we derive a completeness of 72\%.  For the
closest frame limit set at 24" from the nucleus, and plotted in
Figure~\ref{fig_field_comp} as a filled square, we derive a completeness
of only 48\%.  This suggests that a significant incompleteness exists
in our survey within an arcminute of the nucleus of M81.

\begin{figure}
\plotone{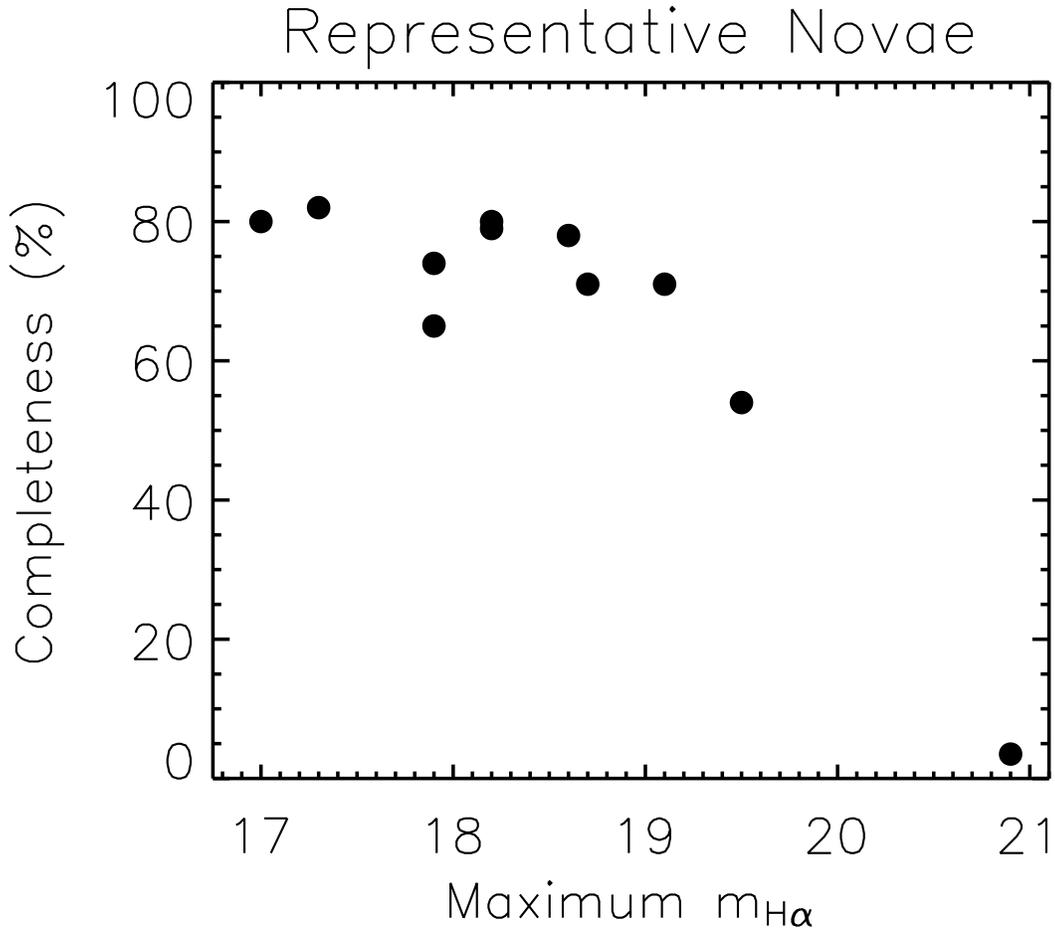}
\caption{The completeness of each of the 11 representative novae as a
function of maximum $m_{H\alpha}$ from the random outburst simulations.
The trend with maximum $m_{H\alpha}$ is obvious, and implies a
survey limit of just fainter than 19 $m_{H\alpha}$.  There is scatter even
for novae with the same maximum $m_{H\alpha}$.  This is due to the
different shapes of the nova light curves and illustrates the difficulty in
assigning a single frame limit to a survey such as ours.
\label{fig_nova_comp}
}
\end{figure}

The completeness as a percentage, averaged over the 15 frame limit sets,
is given for the 11 representative novae in Table~\ref{tab_nova_comp} and
plotted in Figure~\ref{fig_nova_comp}.  The effect of maximum magnitude
on completeness is readily apparent.  This trend is not without scatter,
however, due to the complex interaction between the shape of the light
curve and the depth of the survey as a function of epoch.

\section{The Nova Rate}

The effects noted above demonstrate that assigning a single limiting
magnitude to our survey is overly simplistic.  This fact, combined with
our dense time coverage, indicates that we can improve on the mean nova
lifetime approach of \citet{cia90} for calculating the nova rate for M81.
A raw global nova rate can be obtained simply by dividing the observed
number of novae that erupted during the survey by the time covered.  
Excluding the two novae that erupted before the start of the survey
(Nova 1 and Nova 3) gives $R = 10 / 0.43$yr $=
23$ yr$^{-1}$ and is obviously a hard lower limit on the M81 nova rate.
This rate also requires no correction for partial spatial coverage
since we are covering the entire galaxy.  For comparison, \citet{sha00}
report a nova rate for M81 of $R = 24\pm8$ yr$^{-1}$, based on 15 novae
discovered over a three year period, and refer to an unpublished study
whose preliminary results indicate that the nova rate in M81 may be
somewhat lower than this.

We can adjust our raw rate by the completenesses shown in
Table~\ref{tab_field_comp}.  Since the novae in fields 2N and 5S were
distributed closer to the nucleus of M81, we must account for the lower
completenesses found.  For field 2N, we use a completeness which is the
average of the two positions measured, or 64\%.  For field 5S, we also
average the completeness measurements to get a 82\% average completeness.
Dividing the number of novae that erupted in each field by the appropriate
completeness fraction and adding, we derive a completeness corrected
number of novae during our 5 month survey of 13.  Using this corrected
number of novae, we find a global M81 nova rate of $R = 13 / 0.43$yr $=
30$ yr$^{-1}$.  This is still a conservative lower limit because by
averaging the completenesses in the two Bulge fields, we are assuming
that the number of novae at each measurement location is the same.
Figure~\ref{fig_pos} shows that the novae are more likely to come from
the regions of lower completeness nearer to the nucleus of M81 and hence
the true completeness correction is probably larger than what we have
used here.

\subsection{The Monte Carlo Approach}

\citet{sha01} describe a Monte Carlo technique which uses the maximum
magnitudes and decline rates of the novae in their Table~11 and their
survey faint limit to find the most probable nova rate in their survey
region.  We used the 11 representative nova light curves described above
combined with our frame limits in a similar Monte Carlo experiment to
derive nova rates for each of our fields.

This technique makes many independent estimates of the observed nova rate
in the given field as a function of the true nova rate $[N_{obs}(N_t)]$.
For a given trial estimate of $N_t$, the true rate, we choose a random
set of light curves and outburst times and use the frame limits to
calculate the number of observed novae, using the candidate criteria
described above.  We repeat this 10$^5$ times and record how many times we
recover the number of nova candidates actually observed for that field.
The estimate of the true nova rate $N_t$ is then incremented and the
process is repeated.  This produces a probability distribution for $N_t$
in the given field.  The best estimate for $N_t$ is that which corresponds
to the peak of this distribution.

\begin{figure}
\plotone{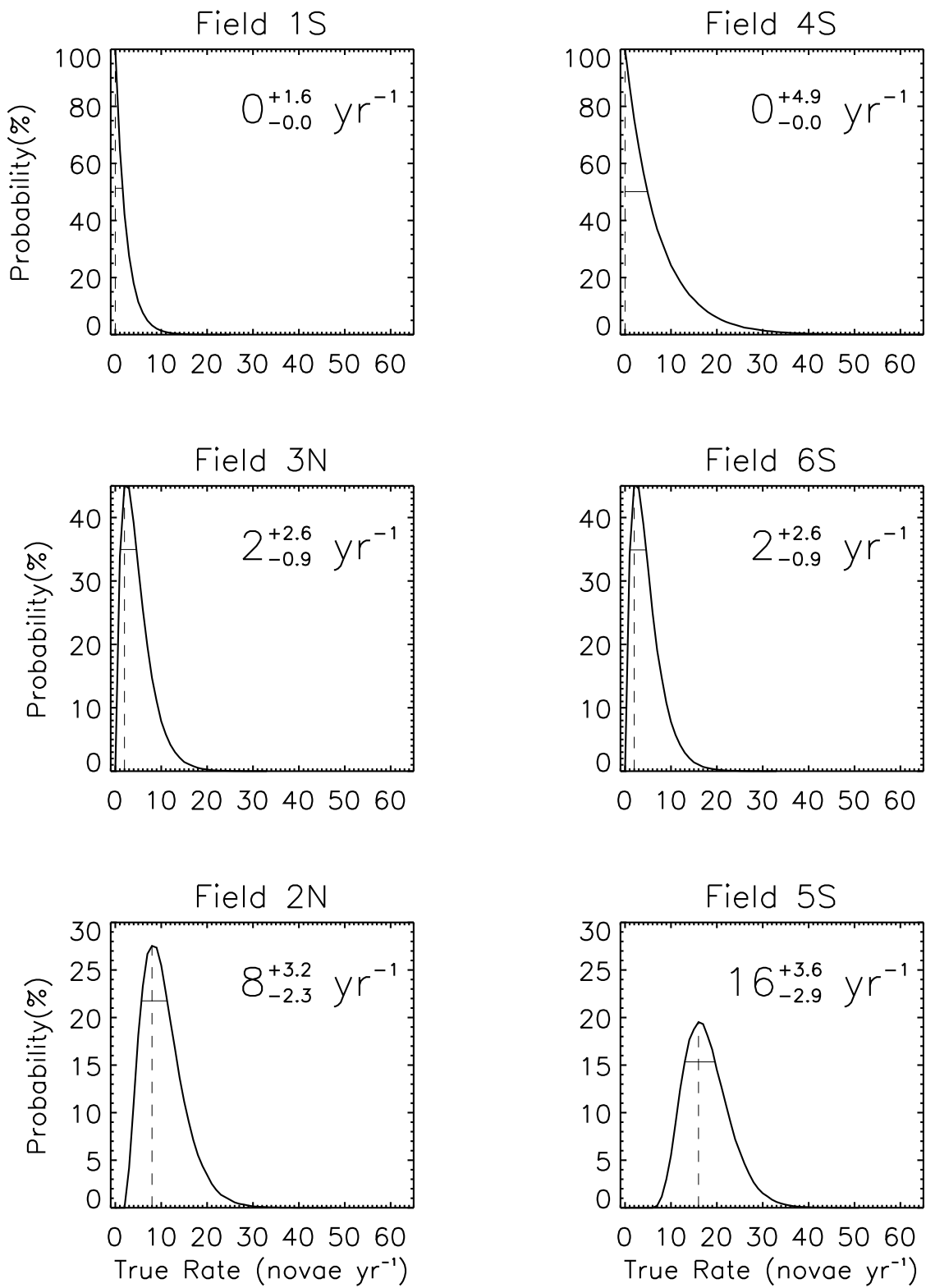}
\caption{Monte Carlo probability distributions of the true nova rate, $N_t$,
for 6 of the 12 M81 fields.  The dashed vertical lines show the locations of
the most probable $N_t$.  The solid horizontal lines show the
limits encompassing half the probability and define the errors for each
estimate of $N_t$.
\label{fig_mc}
}
\end{figure}

Figure~\ref{fig_mc} shows the probability distributions for an interesting
subset of fields.  The two distributions at the top of the figure, for
Field 1S and Field 4S, show the difference between two disk fields with
no observed novae, but very different time coverage and depth.  This is
reflected in the widths of the distributions and consequently the errors
on the estimated true nova rate.  The middle two distributions are for
fields 3N and 6S, having one nova observed in each.  The bottom two
distributions are for the two fields that had the bulk of the observed
novae with 3 for Field 2N and 7 for Field 5S.

Table~\ref{tab_nova_rate} presents the results of the simulations for
each of 15 sets of frame limits.  To derive a global nova rate from these
data, we add up the nova rates in all the fields.  Without accounting
for the incompleteness near the center of M81, i.e., using the frame
limit sets derived at the field centers of the two bulge fields, we
find a global nova rate of 28$^{+10}_{-4}$ yr$^{-1}$ which is in good
agreement with our corrected nova rate of 30 yr$^{-1}$.  If we include the
rates from the frame limit sets closer to the nucleus of M81, as we did
for the completeness, we get a higher nova rate.  Averaging the two rates
for Field 2N we get a rate of 11$^{+7}_{-5}$ yr$^{-1}$.  Averaging the
three rates for Field 5S we get 18$^{+7}_{-6}$ yr$^{-1}$.  Adding these
rates to the nova rates for the other fields, we derive a Monte Carlo,
completeness-corrected nova rate for M81 of 33$^{+13}_{-8}$ yr$^{-1}$.
We point out again that this is a conservative estimate because we have
not accounted for the distribution of novae in our simplistic averaging
of nova rates for fields 2N and 5S.

\subsection{The Luminosity Specific Nova Rate}

The 2-micron All Sky Survey (2MASS) offers a uniform infrared photometric
system for normalizing nova rates from different galaxies to their
underlying stellar luminosity. This will remove one of the major
uncertainties in the study of how the luminosity specific nova rate
(LSNR) in a galaxy varies with Hubble type.  The Large Galaxy Atlas of
the 2MASS \citep{jar03} gives a K-band integrated magnitude for M81 of
3.831$\pm$0.018.  Using this we derive a K-band luminosity for M81 of $L_K
= 8.34\pm0.14 \times 10^{10} L_{\odot,K}$.  We derive a LSNR of $\rho_k =
3.96^{+1.8}_{-1.1}$ yr$^{-1} [10^{10} L_{\odot,K}]^{-1}$ from our nova
rate of 33$^{+13}_{-8}$ yr$^{-1}$, from the Monte Carlo experiment.

This LSNR moves the position of the data point for M81 in Figure~6 from
\citet{sha00} up from $\rho_k = 1.80\pm0.71$.  This does not change the
conclusion they drew from it, namely that no correlation exists between
$\rho_k$ and galaxy $B - K$ color.  There are many systematic and random
errors that could mask such a correlation.  The normalization of nova
rate to infrared luminosity using 2MASS should reduce the scatter
in this diagram caused by non-uniform infrared galaxy magnitudes.
We maintain that the biggest source of error in this diagram is that
nova rates are systematically underestimated for the inner regions of
galaxies where most of the novae may well occur.  This figure could
change substantially as global nova rates are improved and normalized to
a uniform measurement of stellar luminosity for these galaxies.  We now
show that novae probably do appear preferentially in the bulge of M81.

\section{The Spatial Distribution}

\begin{figure}
\plotone{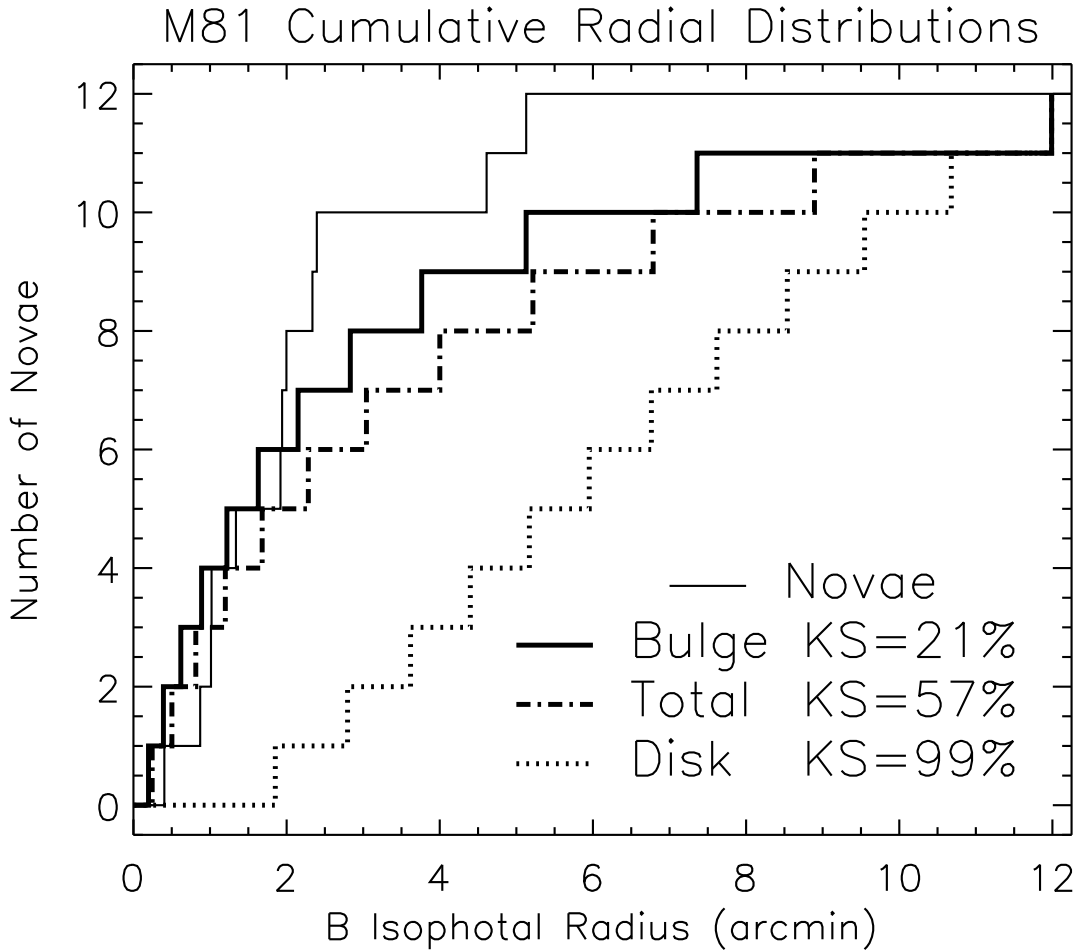}
\caption{The cumulative radial distribution of novae compared with the
bulge/disk decomposition of \citet{sim86} in B-band light.  Using their
parameters we calculated the K-S statistic to test how well we can rule out
the hypotheses that the radial distribution of the novae, indicated by the 
solid thin line, differ from the radial distributions of the various 
components of M81.  The bold solid line represents the bulge light and can 
be ruled out at only the 21\% confidence level.  The total light shown by 
the bold dot-dashed line can be ruled out at the 57\% confidence level.  
The exponential disk, shown by the bold dotted line can be excluded with 
a confidence of 99\%.  While the bulge of M81 does the best, its high 
KS statistic may be due to incompleteness of nova detection in 
the inner part of the galaxy.
\label{fig_ldist}
}
\end{figure}

We do not yet have enough novae in M81 to perform a reliable maximum
likelihood decomposition of bulge and disk novae as did \citet{cia87}
for M31.  We can, however, do a simple test comparing the distribution of
light and novae.  We used the bulge/disk decomposition of \citet{sim86},
Table 4, to calculate the cumulative radial distributions of the bulge,
the disk, and the total light of M81.  We then compared these with the
cumulative radial distribution of the novae.   We used the Kolmogorov
- Smirnov (KS) test to determine at what confidence level we can rule
out the hypotheses that the nova distribution is identical to the three
different light distributions.  Figure~\ref{fig_ldist} shows the results
of this comparison.

The "best match" with the nova distribution of the three is the
bulge light distribution which can only be ruled out at the 21\%
confidence level.  The total light fares considerably worse by over
a factor of two and can be ruled out at the 57\% confidence level.
The disk light distribution is the worst match and can be ruled out at
the 99\% confidence level.  Even though the bulge light does the best,
the confidence level of 21\% is still high.  This is most likely due to
the incompleteness in the central regions of the galaxy (see
Figure~\ref{fig_field_comp}).

Another diagnostic for the bulge-to-disk ratio of novae was described
in \citet{hat97}.  They used a model for dust in M31 and predicted its
effect on the distribution of novae in the bulge region.  They state
that if novae arise primarily in the bulge, a large asymmetry in their
distribution would be apparent because the dust in the disk obscures the
bulge novae behind the disk.  In this case there would be fewer novae
seen on one side of the bulge than on the other.  They use the major
axis of M31 as the dividing line between novae on top of the disk and
novae on the bottom and compared the number of novae below the disk,
on the bottom of the bulge, with the number of novae in the bulge above
the disk.  This ratio is called the bottom-to-top ratio (the BTR).
If disk novae predominate, the asymmetry is much less pronounced across
this line since the fraction of novae obscured by the disk is the same
on each side of the major axis.  Using their dust model, \citet{hat97}
calculated the BTR for two different scenarios for M31: one with a
bulge-to-disk nova ratio of 9 which produced a BTR of 0.33, and one with
a bulge-to-disk nova ratio of 1/2 which produced a BTR of 0.63.

If we look at our sample of novae in M81, we can calculate the same BTR
diagnostic.  M81 clearly has dust in the disk that extends into the bulge
\citep{jac89}.  Its inclination of $60\degpnt4$ \citep{sha99} is not that
different from the inclination of $77\degpnt0$ for M31 \citep{cia87}.
By examining the image of M81 from the Hubble Atlas \citep{san61}, one
can see that the top of the bulge is on the north-east side of the major
axis line.  If we say that the 10 novae within 2\minpnt5 of the nucleus of
M81 are the apparent bulge novae, we see that 2 of these novae are south
of the major axis line (on the bottom) and 8 of these novae are north of
it (on the top) giving a BTR of 2/8 $=$ 0.25.  This BTR is much closer
to the model for M31 from \citet{hat97} with a bulge-to-disk nova ratio
of 9 and clearly rules out the scenario with a bulge-to-disk nova ratio
of 1/2.  One thing we must consider is that our novae were discovered
using H$\alpha$ light, while the models of \citet{hat97} were based
on novae in M31 discovered in the B-band.  Using the redder H$\alpha$
light should reduce the effect of dust on the discovery rate of novae in
the bulge.  One would expect, for the same bulge-to-disk nova ratio, that
the distribution in H$\alpha$ light would be more symmetric, not less,
than the distribution in the B-band.  In addition, the fact that M81 is
slightly more face on than M31 should also reduce the asymmetry, since
a face on galaxy would show no asymmetry regardless of the bulge-to-disk
nova ratio.  The fact that our distribution is even more asymmetric than
their M31 scenario with a bulge-to-disk nova ratio of 9 argues for an
even higher bulge-to-disk nova ratio in M81.

Unless the disk of M81 has a vastly different distribution of dust than
the disk of M31, one must ask why \citet{hat97} conclude that M31 has a
bulge-to-disk nova ratio close to 1/2 while, using the same assumptions,
we conclude that M81 has a bulge-to-disk nova ratio greater than 9.
One possibility is that \citet{hat97} used data from photographic surveys 
which have
been shown to be incomplete in the inner regions of M31 \citep{cia87}.
Another problem, compared with our survey, is that M31 has never been
surveyed comprehensively.  Thus, calculations of the ratio of disk to
bulge novae must account for differences in disk and bulge discovery
rates from fundamentally different surveys, adding large uncertainties
to the calculation.  A uniform, comprehensive survey, such as presented
here, removes these uncertainties.

The spatial distribution of 
the novae in M81 adds more weight to the idea that novae are associated 
with older stellar populations.  It is important to continue to test this 
idea because, if verified beyond a doubt, it would strongly constrain the 
theory of how these objects form and evolve.  We would have to consider that 
novae take much longer to form than previously thought and may not appear 
at all in young stellar populations.

\section{Bulge versus Disk Novae}

We can examine our novae sample in M81 to test the idea that
bulge novae are preferentially fainter and slower than disk novae.
\citet{del98} found that for novae in the Milky Way, the bulge novae
were spectroscopically distinct, having Fe II lines in the early emission
spectrum, slower expansion velocities, and slower decline rates, while the
disk novae had He and N lines, larger expansion velocities, and faster
decline rates.  Unfortunately, the two unambiguous disk novae reported
here, Novae 3 and 10, were not covered well enough to determine their
maximum magnitudes, although their decline rates are toward the slow end of
the distribution (see Table~\ref{tab_nov_prop}).  The presence of Fe II
lines in the spectrum of Nova 6 \citep{fil03} implies that it could be one
of the slow bulge novae.  The decline rate for this nova is, however, the
fastest one observed in M81 and argues that it may be one of the hybrid
objects that evolve from showing Fe II lines to the faster He/N class and
more properly be a member of the fast/bright novae class \citep{del98}.
Nova 6 is 1\minpnt6 from the nucleus of M81 and could be either a disk or
a bulge nova.  These facts are suggestive, and illustrate the difficulty in
testing this idea conclusively.  Clearly we need more comprehensive light 
curves for both bulge and disk novae in order to provide a more definitive 
test.

\section{Future Work}

Adding to the database of complete H$\alpha$ nova light curves will
improve the accuracy of the nova rates derived with Monte Carlo
simulations as the parent population becomes better sampled.  Once we
accumulate enough novae in both the bulge and the disk of M81, we will
be able to perform a decomposition and derive an accurate bulge-to-disk
nova ratio.  If we have enough complete light curves, we should begin to
detect the differences in the bulge and disk novae one would expect from
the results of \citet{del98}.  We will see if the asymmetry in the bulge 
novae across the major axis line persists.  Adding continuum observations 
for the novae in the database will allow us to directly compare the maximum
magnitude decline rate (MMDR) relation in M81 with the MMDR relation
for M31 and determine a nova distance to M81.  We will also be able to
directly compare continuum and H$\alpha$ rise times.

We are currently upgrading the Calypso Telescope WFCAM 2048x2048
pixel CCD to a 4096x4096 pixel CCD which will quadruple the coverage
area (from 5' to 10' on a side).  The filter used for this study was
clearly too narrow and we have ordered a 75\AA~FWHM H$\alpha$ filter to
allow us to compare our nova brightness distribution with that in M31.
The maximum transmission of the new filter will be $>$ 80\% as compared
with 55\% for the 30\AA~wide filter.  The improved throughput combined
with the high resolution afforded by the Calypso Telescope will provide
a substantial improvement in the completeness close to the nucleus of M81.

Using this upgraded equipment, we will conduct an 8 month survey of M81
in 2003-2004 and include either B or V-band observations of the novae
we discover.  Now that our data reduction pipeline is in place, we will
notify the IAU Circulars when each new nova is discovered and hopefully
obtain independent spectroscopic confirmation of each nova candidate.
This will allow us to produce the most accurate nova rate for any
galaxy known.  We hope to extend this survey to include other members
of the M81 group and members of the Local Group as well.

\section{Conclusions}

1) The raw nova rate for M81 provides a hard lower limit at 23 
yr$^{-1}$.  Using a set of representative nova H$\alpha$ light curves in
random outburst simulations we derive a completeness corrected nova rate of
30 yr$^{-1}$.  Monte Carlo simulations using the same set of light curves
provide our most reliable nova rate for M81 of 33$^{+13}_{-8}$ yr$^{-1}$.
The high nova rate found here for M81, with a survey technique that uses
comprehensive time and spatial coverage, implies that the nova rates
for other galaxies derived from surveys with incomplete spatial coverage
and widely spaced epochs are, at best, rough values and may be serious
underestimates.

2) The LSNR for M81 is $\rho_k = 3.96^{+1.8}_{-1.1}$ yr$^{-1} [10^{10} 
L_{\odot,K}]^{-1}$ using our best nova rate and the 2MASS K-band photometry
for M81.  This raises the LSNR for M81 up by a factor of two from that
published by \citet{sha00}, and implies that the LSNR for other
galaxies could be systematically low by similar or greater amounts. 
A definitive comparison between galaxies of different Hubble type
must await the results of comprehensive nova surveys, such as presented
here.

3) The cumulative radial distribution of the novae matches the bulge light
distribution significantly better than either the total or the disk light
distribution, which is ruled out at a 99\% confidence level.  The BTR value
for M81 of 0.25 derived from the asymmetry in the spatial distribution of the 
apparent bulge novae across the major axis line, implies a 
bulge-to-disk nova ratio for M81 of $>$ 9, according to the models of 
\citet{hat97} for novae in M31.  Both these facts lead to an association of
the novae in M81 with the older bulge stellar population.

4) The disk novae reported here, Nova 3 and 10, have decline rates that
place them in the slow class of novae, but their maximum magnitudes were
not determined.  Nova 6 is most likely a fast hybrid nova, is 1\minpnt6
from the nucleus of M81, but cannot be unambiguously assigned to either the
bulge or the disk.  More comprehensive light curves of both bulge and disk
novae are required to test for differences in their average maximum
magnitudes and decline rates.



\acknowledgments

It is a pleasure to acknowledge the Calypso Observatory Director, Edgar
Smith, for his generous allocation of observing time, his support for,
and continued interest in this project.  We also gratefully acknowledge
the support of our observing team, Elaine Halbedel and Viktor Malnushenko,
whose observing abilities and dedication through numerous trials made
this project possible.  We thank Robin Ciardullo for a comprehensive and
rapid referee's report with excellent suggestions.  We also thank Alan
Shafter for valuable comments on the paper.  MS thanks 
Ren\'{e} Racine for conversations and support regarding novae in M81.

This research has made use of the NASA/ IPAC Infrared Science Archive,
which is operated by the Jet Propulsion Laboratory, California Institute of
Technology, under contract with the National Aeronautics and Space
Administration.
This publication makes use of data products from the Two Micron All Sky
Survey, which is a joint project of the University of Massachusetts and the
Infrared Processing and Analysis Center/California Institute of Technology,
funded by the National Aeronautics and Space Administration and the
National Science Foundation.




\clearpage


\clearpage

\clearpage

\clearpage

\clearpage

\clearpage

\clearpage

\clearpage

\clearpage

\clearpage

\clearpage



\begin{deluxetable}{lccrrrcc}
\tablecaption{Observations\label{tab_obs}}
\tablewidth{0pt}
\tablehead{
\colhead{} & \multicolumn{2}{c}{Center}\\
\colhead{} & \multicolumn{2}{c}{(J2000)} \\
\cline{2-3} \\
\colhead{Field} & \colhead{RA} & \colhead{Dec} &
\colhead{Epochs\tablenotemark{a}} & \colhead{\# Exp.} & \colhead{Hours} &
\colhead{Novae} & \colhead{Bulge/Disk}}

\startdata
1N	&09 56 24.6 &68 58 24	&19	&89	&29.7 &0	&Disk	\\
1S	&09 56 24.1 &68 53 44	&18	&87	&29.0 &0	&Disk	\\
2N	&09 55 31.8 &69 01 34	&18	&82	&27.3 &3	&Bulge	\\
2S	&09 55 30.1 &68 56 52	&5	&23	&7.7  &0	&Disk	\\
3N	&09 54 39.9 &69 03 12	&17	&79	&26.3 &1	&Disk	\\
3S	&09 54 40.6 &68 58 29	&10	&48	&16.0 &0	&Disk	\\
4N	&09 56 25.2 &69 07 35	&18	&80	&26.7 &0	&Disk	\\
4S	&09 56 25.3 &69 02 57	&5	&25	&8.3  &0	&Disk	\\
5N	&09 55 31.1 &69 10 58	&6	&29	&9.7  &0	&Disk	\\
5S\tablenotemark{b}&09 55 31.7 &69 06 14&15/21  &68/94 &22.7/31.3 &7 &Bulge\\
6N	&09 54 39.1 &69 12 32	&8	&36	&12.0 &0	&Disk	\\
6S	&09 54 39.4 &69 07 50	&21	&94	&31.3 &1	&Disk	\\
\enddata
\tablenotetext{a}{Except where noted, epochs range from December 31, 2002 
(JD 2452638.8) to June 6, 2003 (JD 2452796.6).}
\tablenotetext{b}{The additional 15 epochs of field 5S were 
acquired between June 6, 2003 and June 26, 2003 to follow the declines
of Nova 11 and Nova 12 (see text).  No additional novae were found in these 
epochs.}
\end{deluxetable}

\clearpage
\begin{deluxetable}{rcccrrr}
\tablecaption{M81 Nova Positions\label{tab_nov_pos}}
\tablewidth{0pt}
\tablehead{
\colhead{} & \multicolumn{2}{c}{Position} & \colhead{} & \colhead{} & 
	\multicolumn{2}{c}{Nuclear Distance}\\
\colhead{} & \multicolumn{2}{c}{(J2000)} & \colhead{} & \colhead{} & 
	\multicolumn{2}{c}{(arcsec)}\\
\cline{2-3} \cline{6-7} \\
\colhead{Nova} & \colhead{RA} & \colhead{Dec} &
\colhead{Field} & \colhead{Detections} & \colhead{Uncorrected} &
\colhead{Corrected} }

\startdata
1	&09 55 39.2 &69 04 22	&5S	&11	&42	&84\\
2	&09 55 21.5 &69 05 60	&5S	&9	&140	&141\\
3	&09 54 53.6 &69 06 49	&6S	&5	&274	&317\\
4	&09 55 35.3 &69 03 34	&2N	&2	&24	&24\\
5	&09 55 28.5 &69 05 10	&5S	&10	&79	&85\\
6	&09 55 48.5 &69 03 05	&2N	&3	&96	&123\\
7	&09 55 40.5 &69 03 49	&2N	&6	&40	&66\\
8	&09 55 46.7 &69 04 07	&5S	&8	&73	&140\\
9	&09 55 26.5 &69 04 44	&5S	&4	&61	&61\\
10	&09 55 04.4 &69 03 24	&3N	&4	&157	&304\\
11	&09 55 34.8 &69 05 34	&5S	&14	&99	&144\\
12	&09 55 47.4 &69 04 44	&5S	&16	&91	&183\\
\enddata
\end{deluxetable}

\clearpage
\begin{deluxetable}{rrccc}
\tablecaption{M81 Nova Magnitudes\label{tab_nov_phot}}
\tablewidth{280pt}
\tablehead{
 & \colhead{J. D.} & & & \colhead{Pl. Limit} \\
\colhead{Nova} & \colhead{(+2452000)} & \colhead{$m_{H\alpha}$} & 
\colhead{Err($m_{H\alpha}$)} & \colhead{$m_{H\alpha}$} }

\startdata

1	&   641.80  &18.02   &0.17  &18.5\\
	&   644.79  &18.46   &0.17  &18.8\\
	&   653.88  &18.16   &0.13  &18.8\\
	&   668.78  &18.50   &0.11  &18.9\\
	&   679.64  &18.33   &0.15  &18.7\\
	&   689.88  &\nodata  &\nodata&18.3\\
	&   693.76  &18.93   &0.15  &19.0\\
	&   707.60  &18.69   &0.25  &18.7\\
	&   709.78  &18.88   &0.14  &19.3\\
	&   711.89  &19.20   &0.22  &19.2\\
	&   720.78  &\nodata  &\nodata&18.9\\
	&   723.79  &18.88   &0.17  &19.0\\
	&   728.63  &18.97   &0.21  &19.0\\
	&   735.63  &\nodata  &\nodata&18.5\\
	&   753.79  &\nodata  &\nodata&18.7\\
	&   758.79  &\nodata  &\nodata&18.4\\
	&   769.69  &\nodata  &\nodata&19.3\\
\\

2	&   641.80  &18.28   &0.12  &18.9\\
	&   644.79  &18.05   &0.09  &19.3\\
	&   653.88  &18.26   &0.10  &19.5\\
	&   668.78  &18.67   &0.10  &19.7\\
	&   679.64  &18.62   &0.13  &19.3\\
	&   689.88  &19.02   &0.23  &19.0\\
	&   693.76  &19.27   &0.14  &19.7\\
	&   707.60  &19.41   &0.23  &19.5\\
	&   709.78  &19.38   &0.16  &19.7\\
	&   711.89  &\nodata  &\nodata&19.4\\
	&   720.78  &\nodata  &\nodata&19.7\\
\\

3	&   642.76  &18.05   &0.08  &19.5\\
	&   652.76  &18.66   &0.23  &19.5\\
	&   657.74  &\nodata  &\nodata&18.5\\
	&   670.78  &18.87   &0.11  &20.0\\
	&   680.71  &19.37   &0.29  &19.5\\
	&   705.75  &\nodata  &\nodata&19.9\\
	&   707.91  &\nodata  &\nodata&19.5\\
	&   710.68  &\nodata  &\nodata&19.0\\
	&   718.79  &\nodata  &\nodata&19.0\\
	&   721.80  &19.87   &0.33  &20.0\\
	&   724.89  &\nodata  &\nodata&19.5\\
	&   725.61  &\nodata  &\nodata&19.5\\
	&   728.79  &\nodata  &\nodata&20.0\\
\\

4	&   639.84  &\nodata  &\nodata&19.2\\
	&   643.78  &18.45   &0.23  &19.0\\
	&   653.77  &\nodata  &\nodata&18.5\\
	&   668.68  &19.66   &0.54  &19.7\\
	&   671.88  &\nodata  &\nodata&19.5\\
\\

5	&   668.78  &\nodata  &\nodata&19.5\\
	&   679.64  &18.72   &0.15  &19.1\\
	&   689.88  &\nodata  &\nodata&18.7\\
	&   693.76  &18.55   &0.10  &19.5\\
	&   707.60  &18.57   &0.15  &18.8\\
	&   709.78  &18.55   &0.09  &19.5\\
	&   711.89  &18.88   &0.14  &19.2\\
	&   720.78  &18.79   &0.15  &19.4\\
	&   723.79  &18.29   &0.09  &19.4\\
	&   728.63  &18.20   &0.09  &19.2\\
	&   735.63  &18.37   &0.13  &18.7\\
	&   753.79  &18.98   &0.18  &19.0\\
	&   758.79  &\nodata  &\nodata&18.9\\
\\

6	&   671.88  &\nodata  &\nodata&20.2\\
	&   676.90  &(19)\tablenotemark{a}     &\nodata&\nodata\\
	&   679.90  &(17.8)\tablenotemark{a}   &\nodata&\nodata\\
	&   680.90  &(17.9)\tablenotemark{a}   &\nodata&\nodata\\
	&   687.99  &17.34   &0.07  &18.7\\
	&   692.63  &17.37   &0.04  &19.7\\
	&   706.84  &19.20   &0.17  &19.7\\
	&   709.63  &\nodata  &\nodata&18.7\\
	&   711.72  &\nodata  &\nodata&19.5\\
\\

7	&   692.63  &\nodata  &\nodata&19.7\\
	&   706.84  &18.26   &0.12  &19.7\\
	&   709.63  &17.42   &0.08  &18.7\\
	&   711.72  &16.97   &0.04  &19.5\\
	&   720.62  &17.66   &0.09  &19.5\\
	&   723.61  &17.79   &0.13  &19.0\\
	&   734.79  &18.82   &0.14  &20.2\\
	&   758.70  &\nodata  &\nodata&19.3\\
	&   765.65  &\nodata  &\nodata&19.7\\
\\

8	&   693.76  &\nodata  &\nodata&19.5\\
	&   707.60  &17.76   &0.09  &18.8\\
	&   709.78  &17.79   &0.05  &19.5\\
	&   711.89  &17.95   &0.07  &19.2\\
	&   720.78  &18.46   &0.13  &19.4\\
	&   723.79  &18.64   &0.12  &19.4\\
	&   728.63  &18.91   &0.17  &19.2\\
	&   735.63  &\nodata &\nodata&18.7\\
	&   753.79  &\nodata &\nodata&19.0\\
	&   758.79  &\nodata &\nodata&18.9\\
	&   769.69  &\nodata &\nodata&19.5\\
\\

9	&   693.76  &\nodata  &\nodata&19.5\\
	&   707.60  &18.69   &0.20  &18.8\\
	&   709.78  &18.74   &0.12  &19.5\\
	&   711.89  &19.12   &0.19  &19.2\\
	&   720.78  &19.36   &0.29  &19.4\\
	&   723.79  &\nodata  &\nodata&19.4\\
\\

10	&   728.71  &\nodata  &\nodata&19.7\\
	&   735.85  &\nodata  &\nodata&18.5\\
	&   754.77  &18.59   &0.10  &19.2\\
	&   759.65  &18.92   &0.12  &19.7\\
	&   770.69  &18.78   &0.17  &19.5\\
	&   778.74  &\nodata  &\nodata&19.2\\
	&   785.75  &19.68   &0.40  &19.7\\
	&   796.63  &\nodata  &\nodata&18.7\\
\\

11	&   758.79  &\nodata  &\nodata&18.9\\
	&   769.69  &17.61   &0.06  &19.7\\
	&   777.66  &17.47   &0.08  &18.9\\
	&   778.66  &17.59   &0.04  &19.9\\
	&   783.65  &17.98   &0.11  &19.5\\
	&   794.64  &18.23   &0.08  &20.1\\
	&   797.64  &18.47   &0.12  &19.5\\
	&   798.65  &18.90   &0.19  &19.2\\
	&   800.64  &18.21   &0.24  &18.5\\
	&   801.68  &18.64   &0.22  &18.9\\
	&   802.68  &18.61   &0.19  &19.5\\
	&   805.67  &\nodata  &\nodata&19.1\\
	&   807.67  &18.72   &0.17  &19.2\\
	&   808.64  &18.82   &0.20  &19.3\\
	&   809.67  &\nodata  &\nodata&18.5\\
	&   810.65  &\nodata  &\nodata&19.0\\
	&   811.66  &\nodata  &\nodata&18.7\\
	&   812.66  &\nodata  &\nodata&19.1\\
	&   813.66  &18.74   &0.16  &19.3\\
	&   815.67  &18.57   &0.14  &19.3\\
	&   816.67  &\nodata  &\nodata&18.9\\
\\

12	&   783.65  &\nodata  &\nodata&19.5\\
	&   794.64  &18.60   &0.11  &20.1\\
	&   797.64  &18.18   &0.10  &19.5\\
	&   798.65  &18.29   &0.12  &19.2\\
	&   800.64  &18.25   &0.29  &18.5\\
	&   801.68  &18.48   &0.19  &18.9\\
	&   802.68  &18.08   &0.11  &19.5\\
	&   805.67  &17.96   &0.15  &19.1\\
	&   807.67  &17.90   &0.10  &19.2\\
	&   808.64  &17.98   &0.09  &19.3\\
	&   809.67  &17.90   &0.14  &18.5\\
	&   810.65  &18.04   &0.13  &19.0\\
	&   811.66  &17.91   &0.11  &18.7\\
	&   812.66  &18.43   &0.14  &19.1\\
	&   813.66  &18.54   &0.14  &19.3\\
	&   815.67  &18.52   &0.14  &19.3\\
	&   816.67  &18.70   &0.21  &18.9\\

\enddata
\tablenotetext{a}{Nova 6: Unfiltered magnitudes from the KAIT Telescope 
\citep{wei03}.}
\end{deluxetable}

\clearpage
\begin{deluxetable}{rrcrcrr}
\tablecaption{M81 Nova Light Curve Properties\label{tab_nov_prop}}
\tablewidth{0pt}
\tablehead{
\colhead{} & \colhead{} & \colhead{} & \colhead{} & \multicolumn{3}{c}{Decline}\\
\cline{5-7} \\
\colhead{} & \colhead{Max} & \colhead{Baseline} & \colhead{Rise Time} & \colhead{Baseline} &
\colhead{Pts} & \colhead{Rate}\\
\colhead{Nova} & \colhead{($m_{H\alpha}$)} & \colhead{(days)} & \colhead{(days)} &
\colhead{(days)} & \colhead{(N)} & \colhead{($m_{H\alpha}$ day$^{-1}$)} }

\startdata
1	&$<$ 18.0	&87	&\nodata&87	&11	&0.010\\
2	&18.0		&68	&$>$ 3	&65	&8	&0.021\\
3	&$<$ 18.0	&79	&\nodata&79	&5	&0.027\\
4	&$<$ 18.4	&25	&\nodata&25	&2	&0.049\\
5	&18.2		&74	&$>$ 49	&25	&3	&0.031\\
6	&17.3		&19	&$>$ 11	&14	&2	&0.129\\
7	&17.0		&28	&$>$ 5	&23	&4	&0.078\\
8	&$<$ 17.8	&21	&\nodata&21	&6	&0.058\\
9	&$<$ 18.7	&13	&\nodata&13	&4	&0.058\\
10	&$<$ 18.6	&31	&\nodata&31	&4	&0.022\\
11	&17.5		&46	&$>$ 8	&38	&13	&0.036\\
12	&17.9		&22	&$>$ 13	&9	&9	&0.087\\
\enddata
\end{deluxetable}

\clearpage
\begin{deluxetable}{lrrrc}
\tablecaption{Completeness of Frame Limit Sets\label{tab_field_comp}}
\tablewidth{0pt}
\tablehead{
\colhead{} & \multicolumn{2}{c}{Nuclear Distance\tablenotemark{a}} \\
\colhead{} & \multicolumn{2}{c}{(arcsec)}\\
\colhead{} & \multicolumn{2}{c}{\rule[15pt]{130pt}{0.5pt}} &
\colhead{Epochs} & \colhead{Completeness} \\
\colhead{Field} & \colhead{Uncorrected} & \colhead{Corrected} & 
\colhead{(N)} & \colhead{(\%)}
}

\startdata
1N	&430	&444	&19	&76\\
1S	&669	&685	&18	&80\\
2N	&141	&199	&18	&80\\
2N\tablenotemark{b}	&24	&24	&18	&48\\
2S	&423	&591	&5	&41\\
3N	&289	&550	&17	&78\\
3S	&431	&841	&10	&79\\
4N	&355	&716	&18	&78\\
4S	&285	&464	&5	&30\\
5N	&424	&564	&6	&40\\
5S	&140	&182	&21	&89\\
5S\tablenotemark{c}	&79	&85	&21	&85\\
5S\tablenotemark{d}	&42	&84	&21	&72\\
6N	&593	&595	&8	&50\\
6S	&372	&431	&21	&79\\
\enddata
\tablenotetext{a}{This distance is from the field center unless otherwise 
noted.}
\tablenotetext{b}{at position of Nova 4}
\tablenotetext{c}{at position of Nova 5}
\tablenotetext{d}{at position of Nova 1}
\end{deluxetable}

\clearpage
\begin{deluxetable}{lccl}
\tablecaption{Completeness of Representative Novae\label{tab_nova_comp}}
\tablewidth{0pt}
\tablehead{
\colhead{} & \colhead{Max} & \colhead{Completeness} &
\colhead{Decline Rate} \\
\colhead{Nova\tablenotemark{a}} & \colhead{($m_{H\alpha}$)} & \colhead{(\%)} &
\colhead{($m_{H\alpha}$ day$^{-1}$)}
}

\startdata
5		&18.2	&79	&0.031\\
7		&17.0	&80	&0.078\\
12		&17.9	&74	&0.087\\
CFNJS 6		&18.2	&65	&0.080\\
CFNJS 20	&19.8	&54	&0.0044\\
CFNJS 26	&17.6	&82	&0.029\\
CFNJS 31	&18.5	&80	&0.013\\
N1992-07	&19.0	&71	&0.0070\\
N1995-06	&18.9	&78	&0.015\\
N1995-07	&19.4	&71	&0.0090\\
N1995-09	&21.2	&3.5	&0.046\\
\enddata
\tablenotetext{a}{Novae from this work are identified with a number,
while novae from \citet{sha01} have the identifications from their Table~11.
The magnitudes from \citet{sha01} have been corrected by +2.48 magnitudes 
to account for the difference in H$\alpha$ filter widths, and the different 
distances of M31 and M81.}
\end{deluxetable}

\clearpage
\begin{deluxetable}{lrrrrr}
\tablecaption{Field Nova Rates\label{tab_nova_rate}}
\tablewidth{0pt}
\tablehead{
\colhead{} & \multicolumn{2}{c}{Nuclear Distance\tablenotemark{a}} \\
\colhead{} & \multicolumn{2}{c}{(arcsec)}\\
\colhead{} & \multicolumn{2}{c}{\rule[15pt]{130pt}{0.5pt}} &
\colhead{Epochs} & \colhead{Novae Obs.} & \colhead{Nova Rate} \\
\colhead{Field} & \colhead{Uncorrected} & \colhead{Corrected} & 
\colhead{(N)} & \colhead{(N)} & \colhead{(yr$^{-1}$)}
}

\startdata
1N	&430	&444	&19	&0	&0$^{+1.7}_{-0.0}$\\
1S	&669	&685	&18	&0	&0$^{+1.6}_{-0.0}$\\
2N	&141	&199	&18	&3	&8$^{+3.2}_{-2.3}$\\
2N\tablenotemark{b}	&24	&24	&18	&3 &14$^{+5.9}_{-4.4}$\\
2S	&423	&591	&5	&0	&0$^{+3.5}_{-0.0}$\\
3N	&289	&550	&17	&1	&2$^{+2.6}_{-0.9}$\\
3S	&431	&841	&10	&0	&0$^{+1.6}_{-0.0}$\\
4N	&355	&716	&18	&0	&0$^{+1.6}_{-0.0}$\\
4S	&285	&464	&5	&0	&0$^{+4.9}_{-0.0}$\\
5N	&424	&564	&6	&0	&0$^{+3.5}_{-0.0}$\\
5S	&140	&182	&21	&7	&16$^{+3.6}_{-2.9}$\\
5S\tablenotemark{c}	&79	&85	&21	&7 &17$^{+3.7}_{-3.2}$\\
5S\tablenotemark{d}	&42	&84	&21	&7 &20$^{+5.0}_{-3.7}$\\
6N	&593	&595	&8	&0	&0$^{+2.8}_{-0.0}$\\
6S	&372	&431	&21	&1	&2$^{+2.6}_{-0.9}$\\
\enddata
\tablenotetext{a}{This distance is from the field center unless otherwise 
noted.}
\tablenotetext{b}{at position of Nova 4}
\tablenotetext{c}{at position of Nova 5}
\tablenotetext{d}{at position of Nova 1}
\end{deluxetable}


\end{document}